\documentclass[a4paper,11pt]{article}
\pdfoutput=1 

\usepackage{jheppub} 

\usepackage{graphicx}
\usepackage[figurename=Fig.]{caption}
\usepackage[tablename=Tab.]{caption}
\usepackage{color}
\usepackage{float}
\usepackage{amssymb}
\usepackage{amsmath}
\usepackage{mathrsfs}
\usepackage{textcomp}
\usepackage{bbm}
\usepackage{cases}
\usepackage{makecell}
\usepackage{pdflscape}
\usepackage{subfigure}
\usepackage{pdfpages}
\usepackage{makeidx}
\usepackage{bm}
\usepackage{xstring}
\usepackage{lipsum}
\usepackage{slashed}
\usepackage{multicol}
\usepackage{setspace}
\usepackage{booktabs}
\usepackage[table]{xcolor}
\usepackage{multirow}
\usepackage{braket}
\usepackage[separate-uncertainty]{siunitx}
\usepackage[displaymath,mathlines]{lineno}

\newcommand{\e}[1]{\times 10^{#1}}

\newcommand{\eref}[1]{Eq.\,(\ref{#1})}

\newcommand{\vo}{\vec{o}\@ifnextchar{^}{\,}{}}
\def\up{\mathrm}

\renewcommand{\thetable}{\Roman{table}}

\graphicspath{{PDF/}}

\def\slash#1{\setbox0=\hbox{$#1$}           
   \dimen0=\wd0                                 
   \setbox1=\hbox{/} \dimen1=\wd1               
   \ifdim\dimen0>\dimen1                        
      \rlap{\hbox to \dimen0{\hfil/\hfil}}      
      #1                                        
   \else                                        
      \rlap{\hbox to \dimen1{\hfil$#1$\hfil}}   
      /                                         
   \fi}                                         %
\def\sl#1{\setbox0=\hbox{#1} 
  \dimen0=\wd0 
  \rlap{\hbox to \dimen0{\hss/\hss}}%
  #1} 
  
 \usepackage{etoolbox}          

\newcommand*\linenomathpatch[1]{%
  \cspreto{#1}{\linenomath}%
  \cspreto{#1*}{\linenomath}%
  \csappto{end#1}{\endlinenomath}%
  \csappto{end#1*}{\endlinenomath}%
}
\newcommand*\linenomathpatchAMS[1]{%
  \cspreto{#1}{\linenomathAMS}%
  \cspreto{#1*}{\linenomathAMS}%
  \csappto{end#1}{\endlinenomath}%
  \csappto{end#1*}{\endlinenomath}%
}

\expandafter\ifx\linenomath\linenomathWithnumbers
  \let\linenomathAMS\linenomathWithnumbers
  \patchcmd\linenomathAMS{\advance\postdisplaypenalty\linenopenalty}{}{}{}
\else
  \let\linenomathAMS\linenomathNonumbers
\fi

\linenomathpatch{equation}
\linenomathpatchAMS{gather}
\linenomathpatchAMS{multline}
\linenomathpatchAMS{align}
\linenomathpatchAMS{alignat}
\linenomathpatchAMS{flalign}

\makeatletter
\patchcmd{\mmeasure@}{\measuring@true}{
  \measuring@true
  \ifnum-\linenopenaltypar>\interdisplaylinepenalty
    \advance\interdisplaylinepenalty-\linenopenalty
  \fi
  }{}{}
\makeatother

\definecolor{kugreen}{RGB}{50,93,61}
\definecolor{myblue}{RGB}{30,60,200}
\definecolor{myred}{RGB}{200,60,60}

\title{\boldmath A proposal to the ‘$\frac12$ vs. $\frac32$ puzzle’}

\author[1]{Qiang Li,}
\author[2]{Wei Feng}
\author[3,4]{and Guo-Li Wang}

\affiliation[1]{School of Physical Science and Technology, Northwestern Polytechnical University, Xi'an 710072, China}
\affiliation[2]{School of Electronic Engineering, Xidian University, Xi’an, 710071, China.}
\affiliation[3]{Department of Physics, Hebei University, Baoding 071002, China}
\affiliation[4]{Key Laboratory of High-precision Computation and Application of Quantum Field Theory of Hebei Province, Baoding 071002, China}

\emailAdd{liruo@nwpu.edu.cn}
\emailAdd{wfeng@xidian.edu.cn}
\emailAdd{wgl@hbu.edu.cn}

\abstract{
We reconsider the semileptonic decays of $B\to D_1^{(\prime)}l\bar \nu_l$. The previous theoretical calculations predict a significantly smaller rate for the semileptonic decay of $B$ to $D'_1(J_l=\frac{1}{2})$ compared with that to the $D_1(J_l=\frac{3}{2})$, which is not consistent with the current experimental data. This conflict is the so-called  `$\frac{1}{2}$ vs.~$\frac{3}{2}$ puzzle'. In this work, we propose {a simple scheme  to fix this problem}, where we suppose that the strong eigenstates $D_1^{(\prime)}$ do not coincide with the eigenstate of the weak interaction, since no experimental results show the weak and the strong interactions have to share the same eigenstates. 
Within the framework of this tentative scheme, meson $B$ first weakly decays to {the weak eigenstates} $D_{\alpha(\beta)}$  and then the latter are detected as the $D_1^{(\prime)}$ by the strong products $D^*\pi$. {We predict that there exist two new particles} $D_{\alpha(\beta)}$ with $J^P=1^+$ which were not previously identified. {The good performance of the new scheme in describing the experimental data may hint at new symmetry in the weak decays of $B_{q}$ to $1^+$ heavy-light mesons.} To test the scheme proposed here, we suggest an experiment to detect the difference in the invariant mass spectra of $D_1$ reconstructed from the $B$ weak decay and from the strong decay products. 
}


\begin{document} 

\maketitle
\flushbottom

\section{Introduction}

In the heavy quark limit, the $J^P=1^+$ heavy-light mesons, such as $B_1$, $B_{s1}$, $D_{1}$ and $D_{s1}$, contain a doublet, one with the light quark total angular momentum $J_l=\frac{1}{2}$, and the other with $J_l=\frac{3}{2}$. 
 {Since the $\ket{J_l = \frac32}$  state mainly decays through a $D$-wave barrier, it has a narrow width, while the $\ket{J_l = \frac12}$ state usually decays through a $S$-wave way and its width is much broader.}
 {Notice both the two $1^+$ charmed-strange mesons $D_{s1}(2460)$ and $D_{s1}(2536)$ are narrow, which seems contradictory with the above analysis.} However, this may be caused by the low mass of $D_{s1}(2460)$,  {and hence it cannot strongly decay to the  $D^*K$ channel like the $D_{s1}(2536)$}. 
 The $J^P=1^+$ $(c\bar u)$ state with $J_l=\frac{3}{2}$ is usually labeled as the $D_1$, and that with $J_l=\frac{1}{2}$ as $D_1'$.  {The theoretical calculations of the semileptonic decays rates of $B$ to the $\ket{J_l=\frac{1}{2}}$ state give a much smaller value than that of $B$ to the $\ket{J_l=\frac{3}{2}}$ state. However,  this theoretical prediction is not supported by current experimental data.}   {This is the famous  `$\frac{1}{2}$ vs.~$\frac{3}{2}$ puzzle'\,\cite{Bigi1997,Yaouanc2000,Uraltsev2000,Uraltsev2004}}, which is more clearly showed in \autoref{Tab-BF-Semi}. The theoretical predictions of the branching fraction for the decay $B\!\to \!D_1'l\bar \nu_l$ are generally one order less than that for $B\!\to\! D_1l\bar \nu_l$.  {The significant discrepancies between the theoretical predictions and the experimental results} have been discussed in several works\,\cite{Morenas1997,Yaouanc2003,Bigi2007,Bernlochner2012,Klein2015,Dingfelder2016,Matvienko2016,Ricciardi2016}. Most of these previous theoretical results are derived in the heavy quark limit and the
corrections might be large. It is expected that the $1/m_c$ corrections induce a large mixing between $\ket{J_l=\frac32}$ and $\ket{J_l=\frac12}$ state, which could soften this puzzle\,\cite{Klein2015}. However, since the traditional theoretical results for the strong decays of $D_1$ and $D_1'$ are well consistent with the experimental data\,\cite{Godfrey2005,Close2005,ZhongXH2008,WangZH2018}, introducing a large mixing angle would inevitably change the strong decay calculation results. A dilemma arises here. 
In a previous work\,\cite{WangGL2205}, we found the puzzle can not be overcome by adding only relativistic corrections but can be partly explained in a small special range of  {the string parameter $\lambda$. However, the mixing angle used} leads to an inconsistent mass order between $D_1$ and $D_1'$ according to  the latest experimental data for the mass of $D_1(2430)$\,\cite{PDG2022}.
The absence of satisfactory understanding and explanations may suggest the existence of a special scheme in this type of decay.

On the other hand, {the BESIII collaboration} reported the experimental observations of semileptonic decays $D^+\!\to\! \bar K_1(1270)\bar l \nu_l$ in 2019\,\cite{BESIII2019-K1270}, and the $D^0\!\to\! K_1(1270)e^+\nu_e$ in 2021\,\cite{BESIII2021-K1270}, where the measured branching fractions are  {$2.3\times10^{-3}$ and $1.1\times10^{-3}$, respectively}. The semileptonic decays of $D\!\to\! K_1$ are quite similar with the problem we discussed above. From the experimental results, if the branching fractions $D\!\to\!K_1(1400)$ were comparable with the $D\!\to\!K_1(1270)$, it would be possible to detect the semileptonic decays of $D$ to the broader $K_1$ system. This can help us answer the question whether the `$\frac{1}{2}$ vs.~$\frac{3}{2}$ puzzle' just happened accidentally or generally existed in such decays involving the unnatural parity mesons which occupy spin-parity $J^P=1^+,2^-,3^+,\cdots$.
\begin{table}[h!]
\setlength{\tabcolsep}{5pt}
\centering
\caption{Branching fraction ($\e{-3}$) of semileptonic decays $B_{(s)}\!\to\! D^{(\prime)}_{(s)1}l\bar \nu_l$ $(l=e$ or $\mu)$. The results labeled `BS' are calculated according to the traditional method by using the Bethe-Salpeter (BS) wave functions; `Ansatz' are calculated within the new scheme proposed here. The theoretical uncertainties in our results are calculated by varying the mixing angle $(\theta\pm5)^\circ$. Since the PDG data only gives the fraction of the cascade decay $B\to D^{(\prime)}_1(D^{(\prime)}_1\!\to\! \bar D^{*0}\pi^-) l\bar \nu$,  {we assumed the branching fraction $\mathcal{B} (D_1^{(\prime)}\!\to\! D^{*0}\pi^-) ={2}/{3}$}; and $\mathcal{B}\left(B_s\!\to\!D_{s1}l\bar\nu\right)$ is determined by $\left(2.94\cdot\frac{1+0.85}{0.85}\right)\times10^{-3}$. The last line denotes the summation of the branching fractions $\mathcal{B}(B^-\!\to\! D_1l\bar \nu_l)$ and $\mathcal{B}(B^-\!\to\! D'_1l\bar \nu_l)$.}  \label{Tab-BF-Semi}%
    \begin{tabular}{ccccccccccc}
\toprule[2.0pt]
    Decay	      								&Ansatz 							&{BS} 						&\cite{Morenas1997}&\cite{Dong2014}&\cite{Ebert2000} 	&\cite{Segovia2011} &\cite{Faustov2013}	&PDG\,\cite{PDG2018}\\
\midrule[1.2pt]
    $B^-\!\to \!D_1 l\bar \nu_l$		&$5.06^{-0.40}_{+0.38}$		&$7.82^{-0.28}_{+0.16}$ 	&3.0-5.0	& $7.04$   			& $6.3 $  				&$3.85$  				&- 						&$4.54\pm0.3$\\
    $B^-\!\to \!D'_1 l\bar \nu_l$   		&$3.46^{+0.40}_{-0.38}$		&$0.64^{+0.27}_{-0.16}$	&0.0-0.7	& $0.45$			& $0.9 $				&$1.98$				&- 						&$4.05\pm0.9$\\
    $B_s\!\to \!D_{s1} l\bar \nu_l$   	&$6.03^{-0.46}_{+0.43}$		&$8.44^{-0.32}_{+0.20}$  & - 		&	-				& $10.6$				&$4.77$				&$8.4\pm0.9$		&$5.66\pm1.52$\\
    $B_s\!\to \!D'_{s1} l\bar \nu_l$   	&$4.21^{+0.46}_{-0.43}$		&$1.36^{+0.28}_{-0.25}$	& -  		&	-				& $1.8$				&$1.74$-$5.7$		&$1.9\pm0.2$		&-\\
\midrule[1.2pt]
    $B^-\!\to \!D^{(\prime)}_1l\bar \nu_l$ &8.42						&8.46						&3.0-5.7	& $7.49$   			& $7.2 $  				&$5.83$  				&- 						&$8.54$\\
\bottomrule[2.0pt]
    \end{tabular}%
\end{table}%

This paper is organized as follow. In Section\,\ref{Sec-3}, we try to explore the `$\frac{1}{2}$\,vs.\,$\frac{3}{2}$ puzzle' by re-examining the related weak decay process and the final strong products in experimental measurements, and then we give our proposal on the `$\frac{1}{2}$\,vs.\,$\frac{3}{2}$ puzzle' and the related discussions. Finally we give a brief summary and outlook of this work. A brief review of the calculation methods and some numerical details used are all collected in the appendix \ref{Sec-2}.
\section{Revisit the $B\to D^{(\prime)}_1 l\bar \nu_l$ decays} \label{Sec-3}
In this section we try to deal with the `$\frac{1}{2}$\,vs.\,$\frac{3}{2}$ puzzle'. It should be pointed out that the proposed scheme and our main conclusion here are not sensitive or dependent on the specific quark model or calculation methods introduced in the appendix. We start by rechecking the experimental measurements first.

\subsection{The weak and strong eigenstates}

The effective Hamiltonian responsible for the semileptonic decays of $B$ to $1^+$ charmed mesons can be expressed at the hadronic level as
\begin{equation}
\begin{gathered}
H_\up{eff} = \frac{G_F}{\sqrt2} V_{cb}( \bar l \Gamma_\nu \nu_l )\, \up{tr} \left(\bar  D_{\alpha} \Gamma^\nu  B + \bar  D_{\beta} \Gamma^\nu  B\right ),
\end{gathered} 
\end{equation}
where $B$ and $D_{\alpha(\beta)}$ here denote the fields of the corresponding mesons, and will also be used to denote the corresponding mesons without causing confusion; $\Gamma^\nu =\gamma^5(1-\gamma^\nu)$ represents the Dirac structure of weak vertex; $l$ and $\nu_l$ represent the fields of the lepton and the related neutrino. The two weak eigenstates of the $1^+$ charmed mesons are generally denoted as $D_\alpha$ and $D_\beta$.

First we re-examine the previous theoretical calculations and experimental measurements on the decays of $B\to D^{(\prime)}_1l\bar \nu_l$. In these processes, $D_1$ or $D_1'$ is regarded as a particle participating in both the semileptonic weak decays and the strong decays to the $D^{*}\pi$ which are then detected by the detectors to reconstruct the $D^{(\prime)}_1$.  Namely, in the previous studies, the weak and strong decays of the $D$ meson system share the same eigenstates. Here the weak eigenstates $D_{\alpha(\beta)}$ refer to the direct hadron products in the semileptonic decay of $B$ meson, which are the eigenstates of above Hamiltonian; and the strong eigenstates denote the involved charmed mesons which can directly strongly decay to $D^*\pi$.

For the later one, namely, the strong decays to $D^*\pi$, there is no controversy. The angular momentum $J_l$ is
conserved in the infinite mass limit, and hence $\ket{J_l=\frac32}$ and $\ket{J_l=\frac12}$
states instead of $\ket{^1P_1}$ and $\ket{^3P_1}$ are the strong eigenstates of the QCD Hamiltonian in this limit. 
In the non-relativistic and heavy quark limit, with the help of the Clebsch-Gordan coefficients, the states $\ket{J_l=\frac{3}{2}}$ and $\ket{J_l=\frac{1}{2}}$ can be decomposed  in the basis of $\ket{^1P_1}$ and $\ket{^3P_1}$  as\, (see appendix \ref{App-1} for detailed calculations)
\begin{equation}\label{E-D1-1P}
\begin{pmatrix}
\ket{J_l=\frac{3}{2}}\\
\ket{J_l=\frac{1}{2}}
\end{pmatrix}=
\begin{bmatrix}
+\frac{\sqrt{2}}{\sqrt{3}} & \frac{1}{\sqrt{3}}\\
 -\frac{1}{\sqrt{3}}&\frac{\sqrt{2}}{\sqrt{3}} 
\end{bmatrix}
\begin{pmatrix}
\ket{^1P_1}\\
\ket{^3P_1}
\end{pmatrix}
=
\begin{bmatrix}
\cos\alpha& -\sin \alpha\\
\sin\alpha    & \quad\cos\alpha
\end{bmatrix}
\begin{pmatrix}
\ket{^1P_1}\\
\ket{^3P_1}
\end{pmatrix},
\end{equation} 
which just corresponds to a counter-clock rotation with rotation angle $\alpha=-35.3^\circ$. 
The most recent data shows that $M_{D_1'}=2.412\,\si{GeV}$\,\cite{PDG2022} is slightly lighter than $M_{D_1}=2.421\,\si{GeV}$, which is just opposite with the previous situation. Then from the above equation, it is easy to see that the $\ket{^1P_1}$ state should correspond to the higher mass state compared with the $\ket{^3P_1}$ state. 
The recent data is also consistent with the relevant mass relationship in charmonia and bottomonia systems, namely, $h_{c(b)}(1P)$ is heavier than $\chi_{c(b)1}(1P)$.
In experiments, $D_1$ and $D_1'$ are reconstructed in the $D^*\pi$ invariant mass spectrum as the strong decay eigenstates. Then throughout this work, we take the states $\ket{J_l=\frac32}$ and $\ket{J_l=\frac12}$ as the two strong eigenstates $D_1$ and $D_1'$, respectively, which is also supported by the consistence between the previous  theoretical calculations and the experimental data for the strong decays of $D_1^{(\prime)}$\,\cite{PDG2016,Eichten1993,Godfrey2005,Close2005,ZhongXH2008,WangZH2018}. Namely, even if $D_1$ and $D_1'$ were the mixing states of $\ket{J_l=\frac32}$ and $\ket{J_l=\frac12}$, the mixing effects should be quite small and would not influence the main discussion in this work. 

However, for the former one, namely, the semileptonic weak decays of $B$, we cannot ensure that if the $B$ weakly decays to the $D_1$ and $D_1'$ directly or it first decays to some other states which are the mixtures of $D_1$ and $D_1'$, and the latter two are just the final states we detected. This feature reminds us to review a very similar example, the neutral kaons $\bar K^0$, $K^0$, and $K^0_\up{L}$, $K^0_\up{S}$. The neutral kaons are typically produced by the strong interactions as the strong eigenstates $\bar K^0$ and $K^0$, which are the superposition states as\,\cite{PDG2020}
\begin{equation} \label{E-U-K}
\begin{pmatrix} \ket{\bar K_0} \\  \ket{K_0} \end{pmatrix}
=\frac1{\sqrt2}
\begin{bmatrix} 1 & -1\\ 1 & ~~\,1\end{bmatrix}
\begin{pmatrix} \ket{K^0_\up{L}}\\ \ket{K^0_\up{S}} \end{pmatrix}.
\end{equation}
Then these neutral kaons decay by the weak interactions as the weak eigenstates $K^0_\up{L}$ and $K^0_\up{S}$ with different lifetimes, where the $CP$ is conserved. Then if we detect a $K^0$ or $\bar K^0$, we will have a probability of $50\%$ to find a $K^0_\up{L}$ and another probability of $50\%$ to find a $K^0_\up{S}$. Now if the similar case happens in the $1^+$ charmed mesons, namely, the weak and strong eigenstates do not coincide with each other,  it may be responsible for the `$\frac12$\,vs.\,$\frac32$ puzzle'. 

The neutral kaons hint us to introduce the weak eigenstates $D_\alpha $ and $D_\beta$ which can be generally expressed as the mixtures of the strong eigenstates $D_1$ and $D_1'$, 
\begin{equation} \label{E-U}
\begin{pmatrix} D_\alpha \\ D_\beta \end{pmatrix}
=
\begin{bmatrix} \cos\theta & -\sin\theta\\ \sin\theta & ~\cos\theta\end{bmatrix}
\begin{pmatrix} D_1\\ D_1' \end{pmatrix},
~~~\up{or}~~~
\begin{pmatrix} D_1\\ D_1' \end{pmatrix}
=
\begin{bmatrix} \cos\theta & \sin\theta \\ -\sin\theta & \cos\theta\end{bmatrix}
\begin{pmatrix} D_\alpha \\ D_\beta  \end{pmatrix},
\end{equation}
where {$D_1$ and $D_1'$} decay by the strong interaction as the strong eigenstates where the $J_l$ is conserved under the heavy quark spin symmetry,  similar with $K^0_\up{L}$ and $K^0_\up{S}$ conserving the $CP$.  The symbol $\theta$ denotes the corresponding mixing angle between the weak and the strong eigenstates.
Then if the weak and the strong decays share the same eigenstates, we have $\theta=0$, which is trivial and the standard treatment to this problem, but not established in experiments. In general, the weak eigenstate $D_{\alpha(\beta)}$ may be different from the strong eigenstate $D_1^{(\prime)}$. 
Then the assumed
$D_{\alpha(\beta)}$ would be a more general description to the $1^+$ charmed meson involved in the semileptonic decay of $B$. Notice this treatment in Eq.\,(\ref{E-U}) can naturally recover the standard calculation when $\theta=0$, since then the  $D_{\alpha}$ and $D_\beta$ are exactly the same with the traditional $D_1$ and $D_1'$ respectively. In this work, we are trying to discuss whether there is any possibility that the mixing angle $\theta$ is not equal to 0. Comparing Eq.\,(\ref{E-U}) with Eq.\,(\ref{E-U-K}), $D_{\alpha(\beta)}$ just corresponds to the $\bar K^0(K^0)$ while $D_1^{(\prime)}$ corresponds to the $K^0_{\up{L(S)}}$. {The only difference is that kaons are generated through strong interactions, but they undergo weak decays, while the charmed mesons are produced via weak interactions, but they undergo strong decays.}

Let's examine the actual effects in the measurement of branching fraction between the proposal here and the traditional treatment more detailed. The semileptonic weak decay widths of $B$$\to$$ D_{\alpha (\beta)} l\bar \nu_l$ are  expressed as
\begin{equation}\label{E-Bab}
\begin{gathered}
\Gamma({D_\alpha})=|\mathcal{A}(B\to D_\alpha l \bar\nu_l)|^2,\\
\Gamma(D_\beta)=|\mathcal{A}(B\to D_\beta l \bar\nu_l)|^2,
\end{gathered}
\end{equation}
where $\mathcal{A}$ denotes the corresponding decay amplitude; and the universal phase space integral is omitted for simplicity. {In any one process of above decays, $B$ either directly decays to a $D_\alpha$ or a $D_\beta$ but not the superposition state of the two. Since the weak eigenstates $D_{\alpha(\beta)}$ are assumed to be the real physical states, the two processes are hence different and distinguishable. We can determine which one is actually taken from the lineshapes of the invariant mass spectrum of the produced charmed meson, for the masses and widths of $D_\alpha$ and $D_\beta$ are different. On the other hand, what we really detected in experiments are $D_1$ and $D_1'$, or in fact, their strong decay products. $D_\alpha$ and $D_\beta$ are related to $D_1$ and $D_1'$ by Eq.\,(\ref{E-U}).} Then combining \eref{E-U} and \eref{E-Bab}, we get the widths for $B$ decaying to $D_1$ and $D_1'$,
\begin{equation}
\begin{aligned} \label{E-B-D1}
\Gamma(D_1)&=c^2 \Gamma (D_\alpha) + s^2 \Gamma (D_\beta),\\
\Gamma(D_1')&=s^2 \Gamma (D_\alpha) + c^2 \Gamma(D_\beta).
\end{aligned}
\end{equation}
where $c(s)$ denote the $\cos\theta\,(\sin\theta)$ for simplicity. Notice here the weak eigenstates $D_{\alpha}$ and $D_{\beta}$ are two different physical states but not the virtual intermediate particles. In a semileptonic decay of $B$, the hadronic product could only be a definite $D_\alpha$ or a definite $D_\beta$. Then if we start with a produced $D_\alpha$, {we will have a probability of $c^2$ to detect a $D_1$ with a long lifetime ($\Gamma_{D_1}\sim31\,\si{MeV}$) and a probability of $s^2$ to detect a $D_1'$ with a quite short lifetime ($\Gamma_{D_1'}\sim314\,\si{MeV}$)}, and vice versa for a $D_\beta$. Then we sum over the decay widths instead of the invariant amplitudes.

On the other hand, in the traditional theoretical calculations, namely, states $D_1$ and $D_1'$ are taken as the direct participants of the weak decays, the corresponding results are 
\begin{equation}
\begin{aligned} \label{E-B-D1-Old}
 \Gamma(B\to {D_1}l\bar \nu_l)&=|\mathcal{A}(D_1)|^2=|c\mathcal{A}(D_\alpha)-s\mathcal{A}(D_\beta)|^2=c^2\Gamma(D_\alpha) + s^2 \Gamma(D_\beta) -2sc \Gamma_\up{In},\\
\Gamma(B\to {D'_1}l\bar \nu_l)&=|\mathcal{A}(D'_1)|^2=|s\mathcal{A}(D_\alpha)+c\mathcal{A}(D_\beta)|^2=s^2\Gamma(D_\alpha) + c^2 \Gamma(D_\beta) +2sc \Gamma_\up{In},
\end{aligned}
\end{equation}
where $\Gamma_\up{In} = \mathcal{A}(D_\alpha) \mathcal{A}^*(D_\beta) +  \mathcal{A}^*(D_\alpha) \mathcal{A}(D_\beta)$ denotes the interference part between $D_\alpha$ and $D_\beta$.
Accordingly, by using the BS wave functions and the Mandelstam formalism, the calculated numerical results are listed in \autoref{Tab-BF-Semi} and labeled as `BS'. It is obvious that, in the traditional calculations, the semileptonic $B$ decays have a substantially smaller rate to the $J_l^{P_l}=\frac{1}{2}^+$ doublet than to the $J_l^{P_l}=\frac{3}{2}^+$ doublet, which is  consistent with other theoretical calculations but contrary to {the experimental data labeled as `PDG' in \autoref{Tab-BF-Semi}}.

Comparing the traditional theoretical results \eref{E-B-D1-Old} with \eref{E-B-D1}, we find that the difference comes from the interference parts, which may be responsible for the `$\frac{1}{2}$\,vs.\,$\frac{3}{2}$ puzzle'.  Any difference between the weak eigenstates $D_{\alpha(\beta)}$ and the strong ones $D_1^{(\prime)}$ would cause this kind of interference.

\subsection{Test of the new scheme}
To test the assumption proposed here, we provide an experimental proposal. Since the assumed weak eigenstate $D_{\alpha(\beta)}$ might be different from the strong eigenstate $D_1$, the invariant mass spectra of the charmed mesons would also be different when  reconstructed from the weak decay and the strong decay. Namely, we can reconstruct the relative narrow $D_1$ from both the weak $B$ decay process and its strong decay products,
\begin{gather}
M^2_\up{w} = (P-p_l-p_\nu)^2,  \\
M^2_\up{s} = (p_{{D^*}} + p_{\pi})^2,
\end{gather}
{
where $M_\up{w}$ and $M_\up{s}$ denote the invariant masses of the long lifetime $D_1$ meson reconstructed from the $B$ weak decays and from its strong decay products, respectively}.
Then the difference (in both the mass peak and width) between $M^2_\up{w}$ and $M^2_\up{s}$ can be used to deny or verify our proposed assumption.
If the assumed physical states $D_{\alpha(\beta)}$ does not exist, the detected properties of $M^2_\up{s}$ and $M^2_\up{w}$ would be exactly the same in experiments. Otherwise, the two new resonances $D_{\alpha(\beta)}$ do exist in the weak decays as the weak eigenstates. Moreover, since both $D_\alpha$ and $D_\beta$ can be detected as $D_1$, there may exist two charmed peak structures in the invariant mass spectrum $M^2_\up{w}$
with the reconstructed final charmed products labeled as $D_1$.
However, since the detection of neutrino is quite difficult in experiments, an alternative choice may be detecting the above difference in the corresponding nonleptonic decay channels of $B$, namely, replacing the lepton pair with a light charged meson. Under the factorization assumptions, the nonleptonic decays would be quite similar with the semileptonic one and then we can expect they share the same physics concerned here.

Besides the `$\frac12$ vs. $\frac32$ puzzle' itself, there are also some other ways to verify our results. It is obvious that, unless the mixing angle $\theta=0$, this kind of discrepancies would generally exist in all weakly decay modes involved the mesons with the unnatural parity $J^P=1^+,~2^-,~\cdots$.  Namely, this kind of discrepancies would happen generally when the weak and strong decay eigenstates are different. The further experimental information on the $B_c$ to $B^{(\prime)}_{(s)1}$ or $D_1^{(\prime)}$ can also test our assumption proposed here. Our scheme in \eref{E-Bab} also predicts that the branching fractions of $B_c$ to the primed $B_{1(s)}$ and $D_1$ are comparable with those to the unprimed ones, while in the traditional calculations the fractions of the primed ones are negligible compared with the unprimed ones.

Also notice that the sum of the two results in \eref{E-B-D1-Old} and \eref{E-B-D1} are equal when ignoring the small difference in phase space, 
\begin{equation}
\Gamma(D_1)+\Gamma(D_1')=\Gamma(D_\alpha)+\Gamma(D_\beta).
\end{equation}
Namely, the traditional calculations can obtain the right results for the total widths of $B$ to $D_1$ and $D_1'$ regardless of whether the weak and strong decays share the same eigenstates.
This can then be used as a first check on our thoughts proposed here.  The sum of branching fractions for $B$ to $D_1$ and $D_1'$  are listed in the last line of  \autoref{Tab-BF-Semi}, which  show a satisfactory consistence between the theoretical predictions and the experimental data.

\subsection{Determination of the mixing angle $\theta$}
\begin{figure}[h!]
\centering
\includegraphics[width = 0.70\textwidth, angle=0]{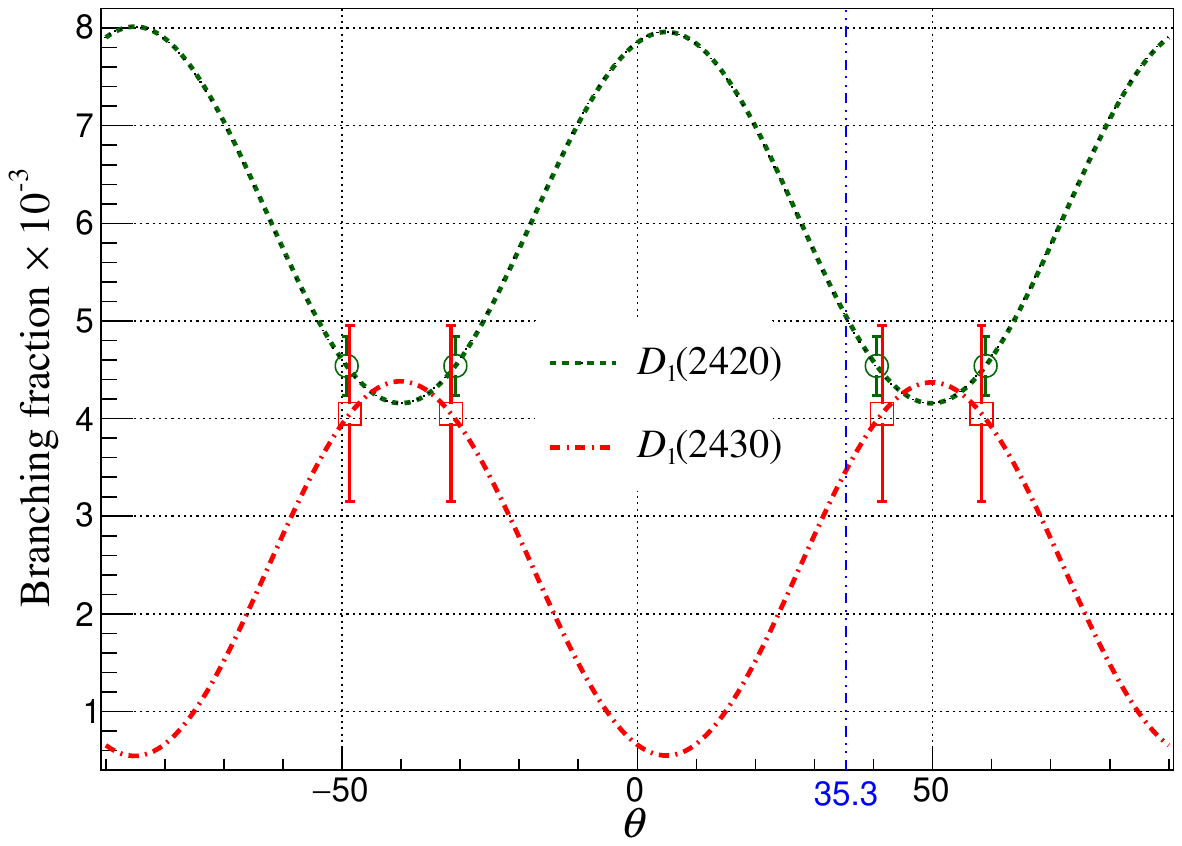}
\caption{Branching fraction of $B\!\to\! D^{(\prime)}_{1}l\bar \nu_l$ versus the mixing angle $\theta$ (defined in Eq.\,(\ref{E-U})). The experimental data are shown by the circle and square respectively. The theoretical calculations can agree well with the experimental data when $\theta=-49.0^\circ$, $-31.3^\circ$, $41.0^\circ$, or $58.7^\circ$, where the latter two correspond to $\vartheta=5.7^\circ$ and $23.4^\circ$ in the basis of $\ket{^1P_1}$ and $\ket{^3P_1}$. }\label{Fig-BFvsMixA2} 
\end{figure}

In fact, we have already given the reason for the `$\frac{1}{2}$\,vs.\,$\frac{3}{2}$ puzzle'.  {In our proposal, this puzzle is caused by the difference between the weak and strong eigenstates.} 
To finally fix the `$\frac{1}{2}$\,vs.\,$\frac{3}{2}$ puzzle', we just calculate the decay widths and then fit to data to obtain the mixing angle $\theta$.  In this work, we calculate the decay branching fractions by Eq.\,(\ref{E-Bab}) combined with Eq.\,(\ref{E-B-D1}) when the mixing angle $\theta$ varies from $-90^\circ$ to $90^\circ$. The obtained results of $B$ to $D_1^{(\prime)}$ are represented in \autoref{Fig-BFvsMixA2}, where we also show the experimental data  {(labeled as `PDG' in \autoref{Fig-BFvsMixA2})} using the circle and square respectively.

{The values of the theta angle that can reproduce the experimental data are} $-49.0^\circ$, $-31.3^\circ$, $41.0^\circ$, and $58.7^\circ$ as shown in \autoref{Fig-BFvsMixA2}. Notice the mixing angle $\theta$ is defined in the basis of $D_1(2420)$ and $D_1(2430)$. Combining Eq.\,(\ref{E-U}) with Eq.\,(\ref{E-D1-1P}), we can also express the weak eigenstates $D_{\alpha(\beta)}$ in the basis $\ket{^1P_1}$ and $\ket{^3P_1}$ as
\begin{equation} \label{E-D-alpha-beta-1P}
\begin{pmatrix} D_\alpha \\ D_\beta \end{pmatrix}
=
\begin{bmatrix} \cos\vartheta & -\sin\vartheta\\ \sin\vartheta & \cos\vartheta\end{bmatrix}
\begin{pmatrix} \ket{^1P_1}\\  \ket{^3P_1} \end{pmatrix},
\end{equation}
where the mixing angle $\vartheta = \theta + \alpha$, and {$\alpha=-35.3^\circ$ obtained in the heavy quark limit (see appendix \label{App-1})}. Then the possible $\vartheta$ is $-84.3^\circ$, $-66.6^\circ$, $5.7^\circ$, or $23.4^\circ$. It is interesting to see that $\vartheta=-84.3^\circ$ and $5.7^\circ$ are in fact equivalent if we interchange the states $D_\alpha$ and $D_\beta$ and then add a global minus sign to $D_\beta$. Also notice a global minus sign to $D_{\alpha(\beta)}$ would not affect the physics. Similarly, the mixing angle $\vartheta=-66.6^\circ$ and $23.4^\circ$ are also equivalent. Namely, the two redundant mixing angles can be eliminated by a proper definition to the two weak eigenstates. Now we let $D_\alpha$ always denote the higher mass one by definition, and then the possible mixing angles which can recover the experimental data are left to be $\vartheta=5.7^\circ$ and $23.4^\circ$. It can be verified that, {under these two mixing angles}, the obtained branching fractions are $4.5\times10^{-3}$ and $4.0\times10^{-3}$ for the semilepontic decays of $B\!\to\! D_1(2420)$ and $B\!\to\! D_1(2430)$ respectively, which agree well with the experimental data. 

Namely, we have fixed the `$\frac12$ vs. $\frac32$ puzzle' by assuming $D_{\alpha(\beta)}$ as the real weak eigenstate and then calculating the corresponding mixing angles. On the other hand, it is easy to see that one of the mixing angles $\vartheta=5.7^\circ$ is quite close to $0^\circ$, which corresponds to the pure $\ket{^1P_1}$ and $\ket{^3P_1}$. Considering the experimental uncertainty and the theoretical errors,  a natural ansatz of $\ket{D_\alpha }$ and $\ket{D_\beta}$ are $\ket{^1P_1}$ and $\ket{^3P_1}$ respectively. Namely, the mixing angle $\vartheta$ in Eq.\,(\ref{E-D-alpha-beta-1P}) may be $0^\circ$.
Roughly speaking, above obtained numerical results further hint us to assume that the states $\ket{^1P_1}$ and $\ket{^3P_1}$, while not the $D_1$ and $D_1'$, might be the real weak eigenstates $D_\alpha$ and $D_\beta$ respectively in the semileptonic decay of a $B$ meson. And $D_{\alpha}$ can then strongly decay to $D^*\pi$ either by $D_1$ with a long lifetime or $D_1'$ with a much shorter lifetime. It should be pointed out that here the assumed particle $D_{\alpha(\beta)}$ is not the virtual intermediate particle but a real physical state. Notice in the traditional studies of the $1^+$ open flavored mesons\,\cite{GI1985,GK1991,Ebert2010}, the $\ket{^3P_1}$ and $\ket{^1P_1}$ are just taken as the theoretical pure states used to obtain the physical $1^+$ mesons by $^1P_1-{^3P_1}$ mixing.

{
Under this ansatz, the obtained branching fractions are $\mathcal{B}(D_\alpha)=6.65\times10^{-3}$ and $\mathcal{B}(D_\beta)=1.86\times10^{-3}$. Then it is easy to find that the branching fractions
of $B^-\to D_1(D_1')l \bar \nu_l$ is $5.1\times10^{-3}$ and $3.5\times10^{-3}$, which is also consistent with the experimental results  and labeled as `Ansatz'  in \autoref{Tab-BF-Semi} for comparison.} From the \autoref{Tab-BF-Semi}, we can see that the new scheme could resolve the `$\frac{1}{2}$\,vs.\,$\frac{3}{2}$ puzzle', and the calculations agree with the data pretty well even without any fine tuning to the mixing angle $\alpha$. The theoretical errors are calculated by varying $\theta$ by $\pm 5^\circ$ to see the dependence on the mixing angle.

\section{Summary and outlook} \label{Sec-4}

In this work, we reconsider the branching fractions of $B$ to the $J^P=1^+$ doublet $D_{1}$ and $D'_{1}$. To resolve the `$\frac{1}{2}$\,vs.\,$\frac{3}{2}$ puzzle', we propose that  $D_1$ and $D_1'$ may not be the eigenstates in such semileptonic weak decays but only be the eigenstates of the strong decays in the final detection,  while the latter case has been well established in both experiments and {theoretical calculations}. {The real weak eigenstates $D_\alpha$ and $D_\beta$ can be expressed as the superposition states of $D_1$ and $D_1'$. The $B$ meson first weakly decays to a $D_\alpha$ ($D_\beta$), and then detected as a $D_1$ or $D_1'$ by the corresponding strong decay products. By fitting to the experimental data,  we found that two mixing angles} $\theta=41.0^\circ$ and $58.7^\circ$ (or equivalently, $\vartheta=5.7^\circ$ and $23.4^\circ$ under the $\ket{^1P_1}$ and $\ket{^3P_1}$ basis, see Eq.\,(\ref{E-D-alpha-beta-1P}) for convention and definition of $\vartheta$), well described the experimental data and then resolve this puzzle.

To test this assumption here, we propose an experiment to detect the difference between $M_\up{w}^2$ and $M_\up{s}^2$, namely, the invariant mass distributions of the long lifetime $D_1$ mesons reconstructed from the $B$ weak decays and from the strong decay products respectively. We also predict that the similar situation would occur in the semileptonic decays of $B_c$ to other $1^+$ heavy-light mesons, such as $B^{(\prime)}_{s1}$, $B^{(\prime)}_{q1}$, and $\bar D^{(\prime)}_{1}$, etc.. Namely, we predicted comparable branching fractions for $B_c$ weakly decaying to the two $1^+$ heavy-light mesons, which can also be used to test the scheme proposed here. Our scheme proposed here may also be tested in the similar processes, such as $D\!\to\! K_1^{(\prime)}l\bar \nu_l$, in the very near future experiments.

\subsection{A further ansatz and discussions}

Besides the above discussions, the obtained numerical results hint us to make an interesting ansatz that the real weak eigenstates $D_\alpha$ and $D_\beta$ may be the states represented by the wave functions $\psi(^1P_1)$ and $\psi(^3P_1)$ respectively. Under this ansatz, the theoretical predictions could also agree with the experimental data pretty well. The ansatz hints that $B$ could only directly weakly decay to the $J^P=1^+$ $(c\bar q)$ systems represented by the wave functions with the definite behaviors (1 or -1) under charge conjugation transformation, but not mixtures of the two, and then the produced states are detected as $D_1$ or $D_1'$. If this ansatz is correct, this phenomena should also appear in the weak decays, $B_c\!\to\! D_1(B_1,B_{s1})l \nu_l$,  $B_s\!\to\!D_{s1}^{(\prime)} l \nu_l$ etc, or to the unnatural parity $J^P=2^-$ mesons, which means we need to reconsider  all the weak decays involving the unnatural parity mesons. 

Within the new scheme, even without knowing the exclusive decay widths of $B$ to the two weak eigenstate $D_\alpha$ and $D_\beta$, we can estimate the corresponding  ratio for the decay width of $B\to D_1l\bar \nu_l$ over that of $B\to D_1'l\bar \nu_l$. From Eq.\,(\ref{E-B-D1}), this ratio can be expressed as
\begin{gather}
R(D_1,D_1') \equiv \frac{\Gamma(B\to D_1l\bar \nu_l)}{\Gamma(B\to D'_1l\bar \nu_l)}
= \frac{1 + \tan^2\theta R(D_\alpha,D_\beta)}{\tan^2\theta +  R(D_\alpha,D_\beta)},
\end{gather} 
where we define the ratio $R(D_\alpha,D_\beta)\equiv \Gamma(B\!\to\! D_\alpha l\bar \nu_l)/\Gamma(B\!\to\! D_\beta l\bar \nu_l)$, and $R(D_\alpha,D_\beta)$ can vary from 0 to $\infty$. Then it is easy to see that the ratio locates in the range of $\tan^2\theta$ to  ${1}/{\tan^2\theta}$, and only depends on the mixing angle $\theta$. In the heavy quark limit, the  mixing angle $\theta$ is predicted to be equal to $\alpha=-35.23^\circ$ (see appendix \ref{App-1}). Then we obtain that the ratio $R(D_1,D_1')$ locates in the range of $\frac12$ to 2, namely, in the order of one, which also well describe the experimental data.

The above discussion is even more applicable to the $J^P=1^+$ bottomed mesons where the heavy quark limit approximation works better than the charmed ones.
Namely, we can also predict that the branching fractions of $B_c$ to the primed $B_{1(s)}$ and $D_1$ are in the same order with the unprimed ones, while in the traditional calculations the fractions of the primed ones are negligible compared with the unprimed ones. For example, within this framework, we predict the branching fraction of $D\!\to\! K_1(1400) \bar l\nu$ are comparable\,($\sim60\%$) with that to the $K_1(1270)$, while in the traditional calculations the former one is one or two orders less than the latter\,\cite{BianLZ2021,ChengHY2017,Khosravi2009}. We also suggest the experiments to measure the branching fraction of $D\!\to\! K_1(1400)\bar l\nu$, which at least has two effects, to see if the  `$\frac{1}{2}$\,vs.\,$\frac{3}{2}$ puzzle'  happens in the $D$ decay, and to check if the weak and strong decays share the same eigenstates.

The good performance of this ansatz hints that here may exist some more deeper physical constraints or symmetry requirements in the weak decays involving $1^+$ heavy-light mesons, which restrict the wave functions of the weak decay eigenstates produced in $B$ mesons must have certain forms. Finally, it should be pointed out that we propose the weak eigenstates $D_{\alpha(\beta)}$ to resolve the `$\frac12$ vs. $\frac32$ puzzle', while the ansatz $\vartheta=0$ is not necessary but interesting and intriguing.

\appendix

\section{Semileptonic decays of $B$ within the Bethe-Salpeter methods}\label{Sec-2}

The detailed numerical calculations in this work are studied within the framework of the instantaneous Bethe-Salpeter\,(BS) methods\,\cite{SB1951,Salpeter1952}, which has already been successfully used to cope with the doubly heavy baryons\,\cite{LiQ2020},  the recently observed exotic pentaquarks\,\cite{XuH2020} and the fully heavy tetraquarks\,\cite{LiQ2021}, and also generally applied to the meson mass spectra\cite{Chang2005A,Chang2010}, the hadronic transitions and decays\cite{WangZ2012A,WangT2013,WangT2013A}. The theoretical calculations from BS methods have achieved satisfactory consistences with the experimental results.
The semileptonic decays of the $B_{(s)}$ to a charmed meson can then be directly calculated by the BS wave functions. Here we briefly review the Bethe-Salpter equation, the corresponding interaction kernel and the relevant wave functions for the two-body meson systems. 

It should be noted that, our proposal used to solve the `$\frac12$\,vs.\,$\frac32$ puzzle' does not depend on the specific calculation methods. The methods introduced here allow us to present a numerical result to complete the topic discussed in this work. 

\subsection{Bethe-Salpeter equation under the instantaneous approximation}
In momentum space, the Bethe-Salpeter equation (BSE) for the bound state of the two-fermion system can be expressed as,
\begin{equation}
\Gamma(P,q)=\int \frac{\up d^4 k}{(2\pi)^4}iK(s)[ S(k_1)\Gamma(P,k) S(-k_2)] ,
\end{equation}
where $\Gamma(P,q)$ is the four-dimensional BS vertex; $P$, the total momentum of the meson; $S(k_1)$ and $S(k_2)$ are the Dirac propagators of the quark and antiquark respectively; $iK(s)$, the interaction kernel with $s=(k-q)$ denoting the exchanged momentum inside the meson; the internal momenta $q$ and $k$ are defined as,
\[q=\alpha_2p_1-\alpha_1p_2,~~~k=\alpha_2k_1-\alpha_1k_2,\]
with $\alpha_i\equiv \frac{m_i}{m_1+m_2}$, and $ m_{1(2)}$ denoting the constituent quark~(anti-quark) mass; $p_1(k_1)$ and $p_2(k_2)$ denote the momenta of the quark and anti-quark respectively. The BS wave function is defined as 
\begin{equation}
\psi(P,q)\equiv S(p_1)\Gamma(P,q) S(-p_2).
\end{equation} 

Under the instantaneous approximation, the interaction kernel does not depend on the time component of $s$. Then the QCD-inspired interaction kernel used in the Coulomb gauge behaves as\,\cite{Chao1992,DingYB1993,DingYB1995,Kim2004},
\begin{equation}
iK(s\,)\simeq i \left[ (2\pi)^3 \delta^3(\vec s\,)\left(  {\lambda}/{a_2}+V_0 \right)- \frac{8\pi \lambda}{(\vec s\,^2+a_2^2)^2}  -\frac{4}{3} \frac{4\pi \alpha_s(\vec s\,)}{\vec s\,^2+a_1^2}\right]\gamma^\alpha \otimes \gamma_\alpha,
\end{equation}
where $\frac{4}{3}$ is the color factor; $a_{1(2)}$ is introduced to avoid the divergence in small momentum transfer zone; the kernel describing the confinement effects is introduced phenomenologically, which is characterized by the the string constant $\lambda$ and the factor $a_2$. The potential used here originates from the famous Cornell potential\,\cite{Eichten1978,Eichten1980}, namely, the one-gluon exchange Coulomb-type potential at short distance and a linear growth confinement one at long distance. In order to incorporate the color screening effects\,\cite{Laermann1986,Born1989} in the linear confinement potential, the potential is modified and taken as the form above. $V_0$ is a free constant fixed by fitting the data. The strong coupling constant $\alpha_s$ has the following form,
\begin{equation}\notag
\alpha_s(\vec s\,)=\frac{12\pi}{(33-2N_f)}\frac{1}{\ln\left(a+ \vec s\,^2/\Lambda^2_{\up{QCD}}\right)},
\end{equation}
where $\Lambda_\up{QCD}$ is the scale of the strong interaction, $N_f=3$, the active flavor number, and $a=e$ is a regulator constant. In this work, we only consider the time component of the kernel\,($\gamma^0\otimes \gamma_0$), for the spatial components of the kernel are always suppressed by a factor $\frac{v}{c}$ in the heavy-light meson systems.

Within the instantaneous kernel, one can further define the Salpeter wave function 
\begin{equation}
\varphi(q_\perp)\equiv i\int\frac{\up dq_P}{2\pi}\psi(q),
\end{equation}
where $q_P=\frac{P\cdot q}{M}$, $q_\perp=q-q_P \frac{P}{M}$ and $M$ is the mass of the bound meson. Then the BSE above can be reduced as the following three-dimensional (Bethe-)Salpeter equation,
\begin{equation}\label{E-SE}
M\varphi(q_\perp)=\ (w_1+w_2) H_1(p_{1\perp})  \varphi(q_\perp) +\frac{1}{2} \left[ H_1(p_{1\perp}) W(q_\perp) -  W(q_\perp) H_2(p_{2\perp})\right],
\end{equation}
where $w_i\equiv (m_i^2-p^2_{i\perp})^{\frac12}$ represents the kinematic energy of the inside fermion, and $m_{1(2)}$ is the constituent mass of the quark\,(anti-quark); 
\begin{gather}
 H_i(p_{i\perp}) \equiv \frac{1}{w_i} H(p_{i\perp}),~~~H(p_{i\perp})=(p^\alpha_{i\perp}\gamma_\alpha+m_i)\gamma^0,
 \end{gather} 
namely $H_i$ is the usual Dirac Hamiltonian $H(p_{i\perp})$ divided by $w_i$; $W(q_\perp)\equiv\gamma^0 \Theta(q_\perp)\gamma_0$ denotes the potential energy part and the three-dimensional BS vertex $\Theta$ behaves as 
\begin{equation}
\Theta (q_\perp) \equiv \int \frac{\up d^3 k_\perp}{(2\pi)^3}K(s_\perp) \varphi(k_\perp).
\end{equation}
The Salpeter wave function $\varphi$ fulfills the following constraint condition,
\begin{equation} \label{E-BS-constraint}
H_1 \varphi(q_\perp) + \varphi(q_\perp) H_2 =0.
\end{equation}
The normalization condition of the Salpeter wave function is expressed as
\begin{equation}
\int \frac{\up d^3 q_\perp}{(2\pi)^3}  \frac{1}{2M} \up{Tr}\,[\varphi^{\dagger}(q_\perp)  H_1\varphi(q_\perp)] =1.
\end{equation}

The numerical values of the model parameters used in this work are just the same with that in the previous calculations \cite{Chang2010,WangZ2012A,WangT2013,WangT2017} and determined by fitting to the corresponding mesons,  namely,
\begin{gather*}
a =e=2.7183,~~ \lambda =0.21\,\si{GeV}^2, ~~ \Lambda_\text{QCD} =0.27\,\si{GeV},  ~~a_1 =a_2=0.06\,\si{GeV};\\
m_u =0.305\,\si{GeV},~~ m_d =0.311\,\si{GeV}, ~~ m_s =0.5\,\si{GeV}, ~~m_c =1.62\,\si{GeV}, ~~m_b=4.96\,\si{GeV}.
\end{gather*}

\subsection{The involved Salpeter wave functions}
The mesons consisting of different flavors do not occupy the definite charge conjugate parity ($C$-parity). Therefore, the wave functions of open flavored mesons with $J^P=1^+$ are usually the mixtures of the wave functions with $J^{PC}=1^{+-}$ and $1^{++}$, which just correspond to the states $\ket{^1P_1}$ and $\ket{^3P_1}$, respectively,  in the non-relativistic situation. Therefore, we will denote the two sub-components of the $1^+$ wave function as $\varphi(^1P_1)$ and $\varphi(^3P_1)$.
According to the properties under space parity and charge conjugation transformations, these two Salpeter wave functions can be constructed as,
\begin{gather}
\varphi(^1P_1)= \frac{q_\perp \!\cdot\! \xi}{|\vec q\,|} \left ( f_1  + f_2  \frac{\slashed P}{M} + f_3 \frac{\slashed q_\perp}{|\vec q\,|} + f_4 \frac{\slashed P \slashed q_\perp}{M|\vec q\,|} \right)\gamma^5,\label{E-1+wave-A}\\
\varphi(^3P_1)= i\frac{\epsilon_{\mu P q_\perp \xi}}{M|\vec q\,|}\gamma^\mu \left(g_1+ g_2 \frac{\slashed P}{M}+  g_3 \frac{\slashed q_\perp}{|\vec q\,|} +  g_4  \frac{\slashed P \slashed q_\perp}{M|\vec q\,|} \right),\label{E-1+wave-B}
\end{gather} 
where $f_{i}$ and $g_i~(i=1,\cdots,4)$ are the radial wave functions; $\epsilon_{\mu P q_\perp \xi}=\epsilon_{\mu \nu \alpha \beta }{P^\nu q^\alpha_\perp \xi^\beta}$ and $\epsilon_{\mu \nu \alpha \beta}$ is the  antisymmetric Levi-Civita tensor; the polarization vector $\xi$ fulfills the following Lorentz condition and completeness relationship
\begin{gather}
P\cdot \xi^{(r)}=0,\\
\sum_r \xi_\mu^{(r)}\xi_\nu^{(r)}=\frac{P_\mu P_\nu}{M^2}-g_{\mu \nu},
\end{gather} 
with $r=0,~\pm 1$ denotes the possible polarization states. The constraint condition \eref{E-BS-constraint} can reduce the independent variables into two for each of the Salpeter wave functions above, namely,
$
f_3=-c_{31}f_1$, $f_4=-c_{42} f_2$, $g_3=c_{31}g_1$, $g_4=c_{42}g_2$,
where
\begin{gather}
c_{31} \equiv  \frac{|\vec q\,|(w_1 - w_2)}{m_1w_2+m_2w_1}, ~~c_{42} \equiv  \frac{|\vec q\,|(w_1 + w_2)}{m_1w_2+m_2w_1}.
\end{gather} 
The coefficient $c_{31}(c_{42})$ is symmetric\,(antisymmetric) under the interchange of the quark and antiquark inside a meson.
Also notice both $\varphi(^3P_1)$ and $\varphi(^1P_1)$ Salpeter wave functions contain the possible $S$- and $D$-wave components besides the dominated $P$ partial waves, which reflects the behaviors of the relativistic wave functions.
Also notice in the non-relativistic representations $\ket{^3P_1}$ and $\ket{^1P_1}$, the former one usually corresponds to the higher mass state.

The initial state $B$ meson is in $J^P=0^-$, and the corresponding Salpeter wave function behaves as\,\cite{Kim2004},
\begin{gather}\label{E-0-wave}
\varphi(0^{-})=\Bigl(h_1+ h_2 \frac{\slashed P}{M}+  h_3 \frac{\slashed q_\perp}{|\vec q\,|} +  h_4  \frac{\slashed P \slashed q_\perp}{M|\vec q\,|} \Bigr) \gamma^5,
\end{gather}  
where the two constrain conditions are $h_3=-c_{31} h_1$ and $h_4=-c_{42}h_2$. 
Solving \eref{E-SE}, the numerical results of the involved wave functions can be obtained.

\subsection{Semileptonic decay widths of $B\to D_\alpha l\bar \nu_l$}
\begin{figure}[h!]
\centering
\includegraphics[width = 0.7\textwidth, angle=0]{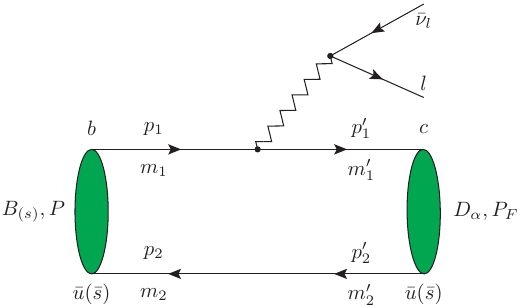}
\caption{Semileptonic decays of $B_{(s)}\!\to\! D_{\alpha}l\bar \nu_l$. $D_\alpha$ denotes the direct charmed hadronic product in the weak decay. $P$ denotes the momentum of $B_{(s)}$, $P_F$ the momentum of $D_{\alpha}$, $p^{(\prime)}_1$ the quark momentum, and $p^{(\prime)}_2$ the anti-quark momentum; $m_{1(2)}$ is the constitute mass of the quark (anti-quark).}\label{Fig-semi-decay} 
\end{figure}
The Feynman diagram\,(tree level) for semileptonic decays of $B_{(s)}$ to a charmed meson is shown in \autoref{Fig-semi-decay}. The invariant amplitude of this process are expressed as
\begin{gather}
\mathcal{A} = \frac{G_F}{\sqrt{2}} V_{cb} \bar u(p_l) \Gamma^\nu v(p_\nu)  \bra{D_\alpha} \bar c \Gamma_\nu b \ket{B},
\end{gather}
where $(\bar c \Gamma_\nu b)$ is the relevant weak current, and $b\,(c)$ denotes the $b\,(c)$-quark field with  $\Gamma^\nu=\gamma^\nu(1-\gamma^5)$; $D_\alpha$ here is used to denote the direct charmed hadronic product with $J^P=1^+$ in the semileptonic weak decay. Notice $D_\alpha$ does not correspond to the strong eigenstate $D_1^{(\prime)}$ naturally, and we will discuss this point more detailed in next section. The hadronic transition amplitude can be generally parameterized by the form factors as,
\begin{gather}
\bra{D_\alpha} \bar c \Gamma_\nu b \ket{B} = \xi_\mu \left(s_1 P^\mu P^\nu +s_2 P^\mu P_F^\nu +s_3 g^{\mu\nu} + i s_4 \epsilon^{\nu \mu PP_F} \right) ,
\end{gather}
where the form factors $s_i\,(i=1,\cdots,4)$ is explicitly dependent on the momentum transfer $(P-P_F)^2$; $\epsilon^{\nu \mu PP_F}=\epsilon^{\nu \mu \alpha \beta} P_\alpha  P_{F\beta}$. On the other hand, the transition matrix element can be expressed by the Salpeter wave function as\,\cite{WangZ2012A,LiQ2016}
\begin{align} \notag
\bra{D_\alpha} \bar c \Gamma_\nu b \ket{B}
&= -i\int \frac{\up{d}^4 q}{(2\pi)^4} \up{Tr} \, \left[\bar \Gamma(P_F,q_F) S(p_1') \Gamma_\nu  S(p_1) \Gamma(P,q) S(-p_2)  \right],
\end{align}
which can then be expressed by the corresponding Salpeter wave functions after performing the  contour integral over $q_P$; the internal momentum $q_F$ in the final state is related to $q$ by $q_{F}=(q+\alpha_2'P_{F}-\alpha_2P)$ with $\alpha_2'=\frac{m_2'}{m_1'+m_2'}$. The form factors $s_i$ can be obtained by finishing the integration above. 
The decay width then can be obtained by performing the integration over the three-body phase space,
\begin{equation}
\Gamma_{B\to D_\alpha l\bar \nu_l}= \frac{1}{2M}  \int \frac{\up{d}^3 \vec P_F}{(2\pi)^3 2E_{\vec P_F}} \frac{\up{d}^3 \vec p_l}{(2\pi)^3 2E_{\vec p_l}}  \frac{\up{d}^3 \vec p_\nu}{(2\pi)^3 2E_{\vec p_\nu}} |\mathcal{A}|^2 (2\pi)^4 \delta^4(P-P_F-p_l-p_\nu),
\end{equation}
where $E_{\vec p_l}=(m_l^2+\vec p\,^2_{\!l})^{\frac12}$ is energy of the charged lepton $l$, and similar for $E_{\vec P_F}$ and $E_{\vec p_\nu}$.

\section{Mixing angle $\alpha$ in the heavy quark limit}\label{App-1}
In the heavy quark limit, the total angular momentum\,($J_l$) of the light quark is conserved and becomes a good quantum number. Then the $J^P=1^+$ heavy-light meson systems can be described either by the light quark total angular momentum $\ket{J_l}$ or the total angular momentum $\ket{^{2S+1}L_J}$, which are related by a rotation. To obtain the corresponding relationships, we use the symbols $s_1$, $s_2$, and $l$ to represent the spin of the heavy quark, light quark, and the orbital angular momentum, respectively. We take the polarization state $\ket{J=1,J_z=1}$ as an example. Then the $\ket{J_l=\frac12}$ state can be expressed by the heavy quark spin $\ket{s_1}$ and the total angular momentum of the light quark $\ket{J_l}$ as,
\begin{align}\label{A-E-1}
\textstyle \ket{J=1,J_z=1;s_1=\frac12,J_l=\frac12} &= \textstyle \ket{\frac12,\frac12}_{s_1}\ket{\frac12,\frac12}_{s_2l}
\end{align}
By using the Clebsch-Gordan\,(CG) coefficients, the state $\ket{\frac12,\frac12}_{s_2l} $ can be decomposed in the basis $\ket{s_2}\ket{l}$ as 
\begin{align}\label{A-E-2}
\textstyle\ket{\frac12,\frac12}_{s_2l} 
&=  \textstyle  -\sqrt{\frac23}\ket{\frac12,-\frac12}_{s_2}\ket{1,+1}_l + \frac{1}{\sqrt3}\ket{\frac12,+\frac12}_{s_2}\ket{1,0}_l,
\end{align}
where the sign convention is kept consistent with that in the PDG\,\cite{PDG2018}.
Inserting the \eref{A-E-2} into the \eref{A-E-1}, and further expressing the spin states $\ket{s_1}\ket{s_2}$ in the coupled representation $\ket{S,S_z}$, we can obtain
\begin{equation}
\begin{aligned}
&\textstyle \ket{J=1,J_z=1;s_1=\frac12,J_l=\frac12} \\
=&  \textstyle  \sqrt{\frac23} \frac{1}{\sqrt2}  \left(\ket{1,1}_{s_1s_2}\ket{1,0}_l - \ket{1,0}_{s_1s_2}\ket{1,1}_l\right)- \frac{1}{\sqrt3}\ket{0,0}_{s_1s_2}\ket{1,1}_l.
\end{aligned}
\end{equation}
Notice that $\frac{1}{\sqrt2}  \left(\ket{1,1}_{s_1s_2}\ket{1,0}_l - \ket{1,0}_{s_1s_2}\ket{1,1}_l\right)$ is just the state $\ket{J=1,J_z=1; S=1}$, and $\ket{0,0}_{s_1s_2}\ket{1,1}_l$ corresponds to the state  $\ket{J=1,J_z=1; S=0}$. Analysis on other polarization states $\ket{J,J_z}=\ket{1,0}$ or $\ket{1,-1}$ can reach the same conclusion.  Namely, we can express the state $\ket{J_l=\frac12}$ in the $\ket{^{2S+1}L_J}$ basis as
\begin{align}
\textstyle \ket{J_l=\frac12} =\textstyle  - \sqrt{\frac{1}{3}}\ket{^1P_1} + \sqrt{\frac23} \ket{^3P_1} .
\end{align}

By a similar calculation, the $\ket{J_l=\frac32}$ state can be expressed as
\begin{align}
\textstyle \ket{J_l=\frac32} =+  \sqrt{\frac23}\ket{^1P_1}+\sqrt{\frac13} \ket{^3P_1}.
\end{align}
Expressing the above results in the matrix form
\begin{equation}
\begin{pmatrix}
\ket{J_l=\frac{3}{2}}\\
\ket{J_l=\frac{1}{2}}
\end{pmatrix}=
\begin{bmatrix}
+\frac{\sqrt{2}}{\sqrt{3}} & \frac{1}{\sqrt{3}}\\
 -\frac{1}{\sqrt{3}}&\frac{\sqrt{2}}{\sqrt{3}} 
\end{bmatrix}
\begin{pmatrix}
\ket{^1P_1}\\
\ket{^3P_1}
\end{pmatrix}
=
\begin{bmatrix}
\cos\alpha& -\sin \alpha\\
\sin\alpha    & \quad\cos\alpha
\end{bmatrix}
\begin{pmatrix}
\ket{^1P_1}\\
\ket{^3P_1}
\end{pmatrix},
\end{equation} 
we can obtain the mixing angle $\alpha=-35.3^\circ$. Note this mixing angle is connected with the convention used here, and if the rotation matrix was written as $ \begin{pmatrix}
\cos\alpha& \sin \alpha\\
-\sin\alpha    & \cos\alpha
\end{pmatrix}$, one should obtain $\alpha=35.3^\circ$.

\acknowledgments
The authors thank Chao-Hsi Chang, Hui-Feng Fu, and Xu-Chang Zheng for helpful discussions. This work is supported by the National Key R\&D Program of China\,(2022YFA1604803), and National Natural Science Foundation of China\,(NSFC) under Grant Nos.\,12005169, 62201438, and 12075073. It is also supported by the Natural Science Basic Research Program of Shaanxi\,(No.\,2021JQ-074), and the Fundamental Research Funds for the Central Universities.

\medskip


\providecommand{\href}[2]{#2}\begingroup\raggedright\endgroup

\end{document}